\documentclass{aa}  

\usepackage{graphicx}
\usepackage{txfonts}
\usepackage{hyperref}
\usepackage[modulo,switch]{lineno}

\hypersetup{
    final=true,
    pageanchor=true,
    colorlinks=true,
    breaklinks=true,
    linkcolor=blue,
    citecolor=blue,
    urlcolor=blue,
    pdfpagemode=UseNone}

\usepackage{natbib}
\bibpunct{(}{)}{;}{a}{}{,}
\usepackage{multirow}
\usepackage{nicefrac}

\begin{document}

   \title{Height variation of magnetic field and plasma flows in isolated bright points}


   \author{Christoph Kuckein}    

   \institute{Leibniz-Institut f\"ur Astrophysik Potsdam (AIP),
              An der Sternwarte 16,14482 Potsdam, Germany\\
              \email{ckuckein@aip.de}}

   \date{Accepted on August 26, 2019}

 
  \abstract
   {}
   {The expansion with height of the solar photospheric magnetic field and the plasma flows 
   is investigated for three isolated bright points (BPs).}
   {The BPs were observed simultaneously with three different instruments attached to the 1.5-meter GREGOR telescope:
   (1) filtergrams of \ion{Ca}{ii} H and blue continuum (4505\,\AA) with the HiFI, 
   (2) imaging spectroscopy of the \ion{Na}{i} D$_2$ line at 5890\,\AA\ with the GFPI, 
   and (3) slit spectropolarimetry in the 1\,$\mu$m spectral range with the GRIS.
   Spectral-line inversions were carried out for the \ion{Si}{i} 10827\,\AA\ Stokes profiles.}
   {Bright points are identified in the \ion{Ca}{ii} H and blue continuum filtergrams. Moreover, they are also detected in the blue
   wing of the \ion{Na}{i} D$_2$ and \ion{Si}{i} 10827\,\AA\ lines, as well as in the \ion{Ca}{i} 10839\,\AA\ line-core images.
   We carried out two studies to validate the expansion of the magnetic field with height. On the one hand, we compare 
   the photospheric Stokes $V$ signals of two different spectral lines that are sensitive to different optical 
   depths (\ion{Ca}{i} vs. \ion{Si}{i}). 
   The area at which the Stokes $V$ signal is significantly large is almost three times larger for the 
   \ion{Si}{i} line -- sensitive to higher layers -- than for the \ion{Ca}{i} one.
   On the other hand, 
   the inferred line-of-sight (LOS) magnetic fields at two optical depths ($\log \tau = -1.0$ vs. $-2.5$) from the 
   \ion{Si}{i} line reveal spatially broader fields in the higher layer, up to 51\% more extensive in one of the BPs. 
   The dynamics of BPs are tracked along the \ion{Na}{i} D$_2$ and \ion{Si}{i} lines.  
   The inferred flows from \ion{Na}{i} D$_2$ Doppler shifts are rather slow in BPs ($\lesssim$1\,km\,s$^{-1}$).
   However, the \ion{Si}{i} line shows intriguing Stokes profiles with important asymmetries. The analysis of these
   profiles unveils the presence of two components, a fast and a slow one, within the same resolution element.  
   The faster one, with a smaller filling factor of $\sim$0.3, exhibits LOS velocities of about 6\,km\,s$^{-1}$. 
   The slower component is slightly blueshifted.}
   {The present work provides observational evidence for the expansion of the magnetic field with height. Moreover, 
   fast flows are likely present in BPs but are sometimes hidden because of observational limitations. 
   }


   \maketitle
%

\section{Introduction}

The immediate atmospheric layer below the solar surface is unstable and dominated by convective motions. 
These instabilities are observed as a granular pattern which consists, on its smallest scales, of convective 
cells -- granules -- and between these cells of dark intergranular lanes.
Oftentimes, these relatively dark lanes show small-scale brightenings in various wavelengths 
such as the molecular bandhead of CH (G-band) \citep{steiner01} or wideband \ion{Ca}{ii} H and K lines. 
These very small features are called bright points (BPs) and their average size is
on the order of 0\farcs35 \citep[e.g.,][]{berger95,bovelet03}. 
The existence of BPs has been known for many decades \citep[e.g.,][]{mehltretter74}, 
although often under different terms. For instance, 
\citet{muller83} called them facular points with a mean lifetime
of 18\,min. Recently, \citet{liu18} investigated the lifetimes of about 1300 isolated BPs
of which 10\% had lifetimes longer than 7\,min while the others had shorter 
lifetimes.

Observations indicate a strong correlation between BPs and magnetic 
flux concentrations \citep[e.g.,][]{viticchi10,yang16}. However, 
some studies report that not all BPs present polarization signals. 
In a statistical study,
\citet{beck07} find that 94\% percent of 447 identified BPs were co-spatial 
with polarization signals above the noise level. In another work, \citet{berger01}
analyzed 720 G-band BPs and find that all were of magnetic nature.
Realistic 3D radiative magnetohydrodynamic 
(MHD) simulations support this correlation \citep[e.g.,][]{shelyag04}.
Yet, caution has to be taken since 
not all small magnetic elements necessarily coincide with G-band or \ion{Ca}{ii} H 
brightenings \citep{dewijn09}.

\begin{figure*}[!t] 
   \resizebox{\hsize}{!}{\includegraphics{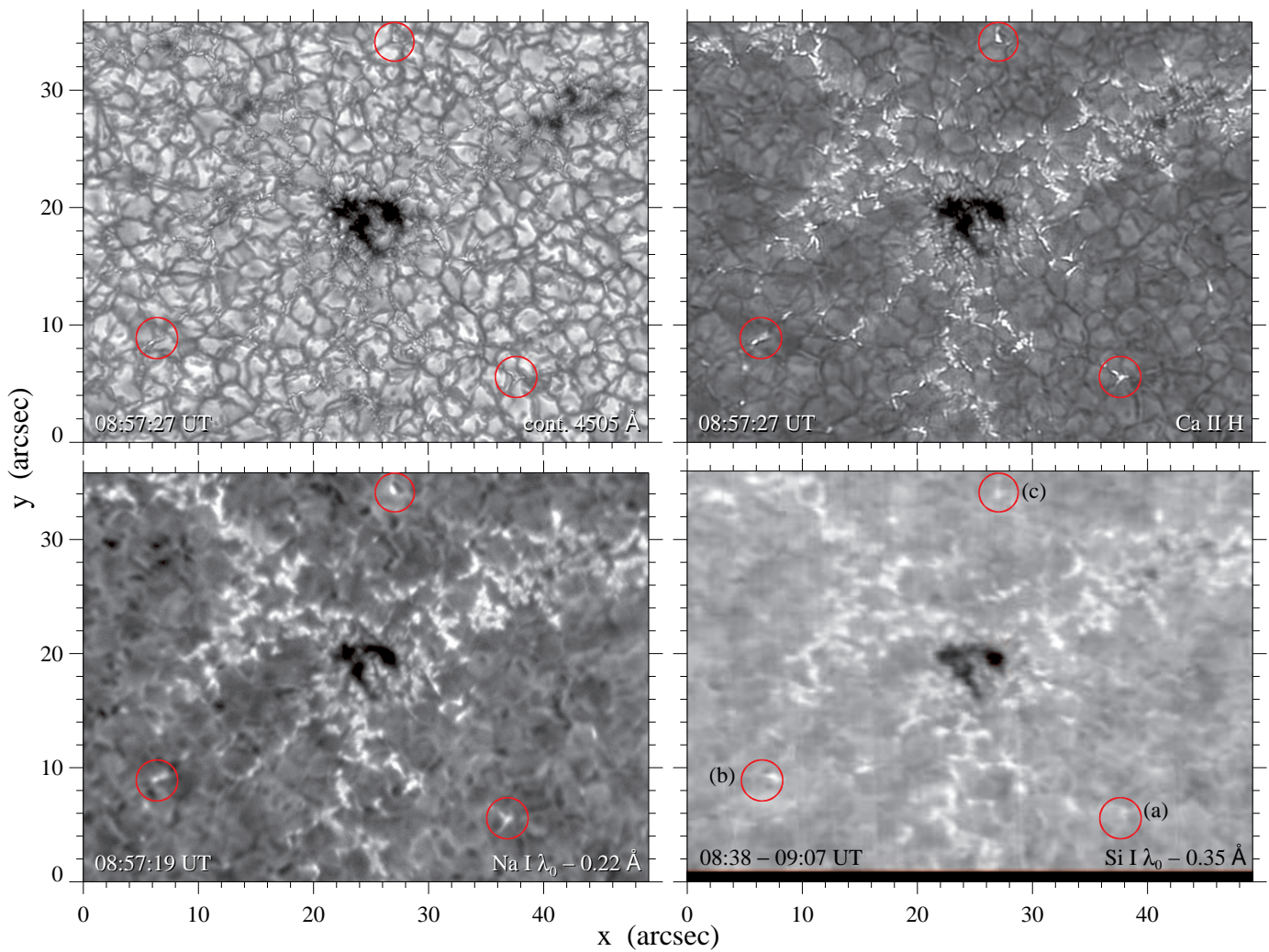}}
      \caption{Overview of small AR NOAA 12708. Starting in the upper left panel and 
      moving clockwise: Speckle-restored blue continuum (4505\,\AA) and \ion{Ca}{ii} H (3968\,\AA)
      images from HiFI, \ion{Si}{i} 10827\,\AA\ slit-reconstructed blue wing ($\lambda_0-0.35$\,\AA) from GRIS, and
      MOMFBD-restored blue-line wing ($\lambda_0-0.22$\,\AA) image from GFPI of the \ion{Na}{i} D$_2$ 5890\,\AA\ line. 
      We note that the 
      image scales from all instruments are different. The black stripe
      in the lower part of the \ion{Si}{i} $\lambda_0-0.35$\,\AA\ image indicates that there are no data available.
      Red circles identified by letters (a)--(c) indicate the isolated BPs under study. 
      It is interesting to note that the BPs in the \ion{Si}{i} $\lambda_0-0.35$\,\AA\ panel do not always match the 
      other instruments because it is a slit-reconstructed image 
      in the range of 08:38--09:07\,UT.}
   \label{Fig:overview}
\end{figure*}

Many studies over the past decades were devoted to the inference of the magnetic 
field strength in BPs. 
From multi-line inversions, \citet{beck07} infer a magnetic field strength 
in the range of 500--1400\,G. The authors treated the filling factor $f$ as a free parameter,
meaning that the area occupied by the magnetic component inside of each pixel can vary. 
This is an important aspect, as BPs are very small features, which are often not fully 
resolved by current telescopes because of insufficient angular resolution. Therefore, if 
a unit filling factor is assumed, the inferred magnetic field strength may change
\citep{criscuoli14}.
\citet{utz13}, assuming $f=1$, report on field strengths of BPs 
which formed a two-component distribution with a clear peak in the kilogauss regime around 1300~G to 1500~G and 
a second tail distribution peaking around a few hundred Gauss. 
The latter might be due to measurement difficulties, such as capturing the BP before or after the convective collapse.
Fields on the order of kilogauss were also found in 3D MHD simulations \citep{riethmuller14}, 
regardless of their presence inside
or outside of active regions \citep{criscuoli14}.
Even strong longitudinal magnetic fields can be inferred
during the formation process of BPs, reaching up to $\sim$2000\,G \citep{nagata08}.

Bright points were observed with very high spatial resolution 
(0\farcs15--0\farcs18) using the SUNRISE balloon-borne solar 
observatory \citep{sunrise}. 
It was found that, when BPs merge, 
an enhancement of the BP brightness is related to an increase 
of the magnetic field strength \citep{requerey15}. The opposite
occurs when BPs fragment. The continuous merging 
and fragmenting of BPs is likely triggered by the evolution of 
granules, which constantly compress and stretch the field lines
\citep{martinezgonzalez11}. Kilogauss fields are then related to
compressed magnetic structures and lower fields to fragmented BPs. 
This explains why a two-component magnetic field
distribution appears in larger statistical studies.  
Moreover, SUNRISE data illustrated large asymmetries of the Stokes $V$
profiles in individual network patches \citep{martinezgonzalez2012}.
These asymmetries varied across the patch, showing that 
the internal structure of the magnetic patch was at least partially
resolved. The results of the spectral-line inversions and the Stokes $V$
asymmetries suggested a geometry with magnetic field lines that expand 
with height. Such a geometry matches with classical models 
\citep[e.g.,][]{spruit76,solanki93}.

Plasma flows related to BPs are found to be directed downward. In the photosphere, 
using separate observations for the \ion{Fe}{i} 5250.5\,\AA\ and \ion{Fe}{i} 6302\,\AA\ lines, 
the Doppler shifts reach up to 
6\,km\,s$^{-1}$ \citep{utz14,nagata08}. Yet, the average values are around 2.4\,km\,s$^{-1}$. 
Further up in the atmosphere, in the upper photosphere or 
lower chromosphere, the line-of-sight (LOS) velocities associated to spectral shifts
of the \ion{Na}{i} D1 line can yield 7\,km\,s$^{-1}$ \citep{jess10}. Hence, a continuous downflow
across several layers of the solar atmosphere seems to take place at BPs. Evidence in this regard
was provided by \citet{fischer09} who found co-temporal large downflows in BPs encoded in two 
photospheric spectral lines, one in the deep photosphere (\ion{Fe}{i} 6302\,\AA) and another 
one in the upper photosphere (\ion{Mg}{i} b2).

This work is based on multi-wavelength observations of BPs acquired with the 1.5-meter 
GREGOR solar telescope \citep{schmidt12}. Thus, we can investigate some BP parameters with height.
Such studies are needed for the modeling of BPs 
\citep[see, e.g.,][]{gent13}. Only by multi-wavelength measurements in various heights of the solar atmosphere 
we will be able to constrain BP models, which are needed for MHD wave propagation and lead to lifelike 
results \citep[e.g.,][]{fedun11}. In this study, three carefully selected isolated BPs 
are analyzed in depth. The spatial resolution is highest in the blue part of the visible spectrum, 
which facilitates studying the small-scale structure of BPs, whereas the near-infrared spectral range 
at 1\,$\mu$m offers a 
spectropolarimetric diagnostic window, including photospheric and chromospheric lines, to determine
the magnetic structure and LOS velocities. 
Sections \ref{Sect:2} and \ref{Sect:3} describe the observations, instrumental setup, 
and data analysis. Section \ref{Sect:4} covers the
results. The results are discussed in Sect. \ref{Sect:5} and the main findings are summarized in the 
Conclusions (\mbox{Sect. \ref{Sect:6}}).

\section{Observations and data reduction} \label{Sect:2}

The target of the observations was a decaying pore, 
prominently surrounded by a network of BPs
and tiny pores, which belonged to active region
NOAA 12708. The active region rotated onto the solar disk on 2018 May 04.
The observations took place a few days later on 2018 May 09. 
The location of the pore was close to disk center, at coordinates $(x,y) = (62\arcsec, -120\arcsec)$,  
where the cosine of the heliocentric angle is $\mu$$=$0.99. The three analyzed BPs are
highlighted with a red circle in Fig. \ref{Fig:overview}. They were chosen because they were isolated, 
that is, far away from the central pore and outside of the densely-packed network of BPs. 
Furthermore, all three BPs were inside of the FOV of all instruments. 
The negative polarity of the magnetic field covered almost the entire FOV except for 
BP (a), which belonged to a small area of positive polarity. 

The instrumental setup at the 1.5-meter GREGOR telescope 
consisted of three different instruments, which were 
recording data simultaneously. The analyzed time range spans between 08:38--09:07\,UT, roughly 30~minutes. 
A detailed description of the used instrumentation follows below. 

\subsection{Spectropolarimetric rasters scans in the near infrared}
The GREGOR Infrared Spectrograph \citep[GRIS,][]{collados12}
acquired full-Stokes spectra in the \ion{Si}{i} 10827\,\AA\ spectral range. 
The number of accumulations per step were ten, with an exposure time of 100 ms
each. The whole map encompassed 300 steps (roughly 30\,min), with a step size and pixel size along 
the slit of 0\farcs135, yielding a field of view (FOV) of 60\arcsec$\times$40\arcsec.
The data was dark and flat-field corrected, as well as polarimetrically calibrated using
the GREGOR polarimetric unit \citep{hoffmann12} and the standard 
procedures \citep{collados99,collados03}. The Stokes profiles were normalized to the average continuum, 
which was determined from different areas across the FOV excluding regions with significant polarization signals. 

The spectral range comprised the photospheric \ion{Ca}{i} 10838.97\,\AA\ and \ion{Si}{i} 10827.09\,\AA\ lines. 
Both lines are Zeeman triplets with an effective Land\'e factor of $g_\mathrm{eff} = 1.5$ \citep{balthasar16}.
Moreover, the spectral range included the chromospheric \ion{He}{i} 10830\,\AA\ triplet. 
The spectral sampling was 18.04\,m\AA\,pixel$^{-1}$. 
The wavelength was calibrated on an absolute scale using the two telluric lines next to the 
\ion{He}{i} 10830\,\AA\ triplet and taking into account orbital motions, solar rotation, and
gravitational redshift. A detailed description of the wavelength calibration is given in the appendices of
\citet{martinez97} and \citet{kuckein12b}.
All reference wavelengths throughout the manuscript were taken from the 
National Institute of Standards and Technology \citep[NIST,][]{NIST}.
Bright points in the near-infrared, within the FOV of GRIS, are well identified in the core of the \ion{Ca}{i} line and 
in the blue wing of the \ion{Si}{i} line at 
$\lambda_0-0.35$\,\AA\ (lower right panel of \mbox{Fig. \ref{Fig:overview}}).

\begin{figure}[!t]
 \centering
 \includegraphics[width=\hsize]{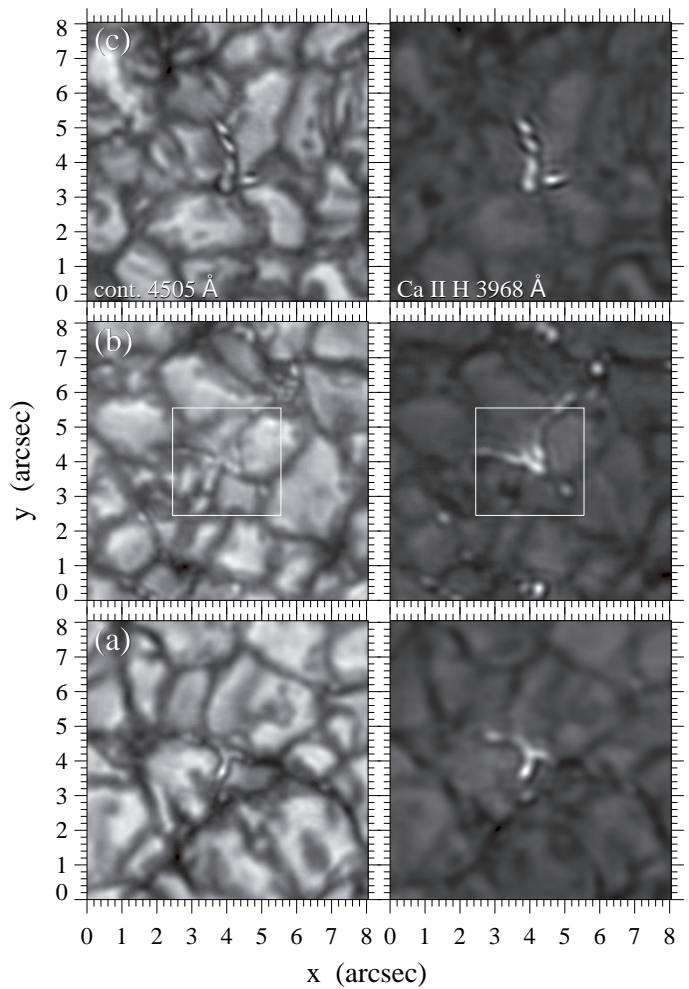}
\caption{Region-of-interest including the three isolated BPs (a)--(c) from Fig. \ref{Fig:overview}. 
  The left column shows blue-continuum images whereas the right
  column shows the corresponding \ion{Ca}{ii} H image. 
  The white box shows the size of the FOV of the inversions shown in Fig. \ref{Fig:inversion_maps}.} 
  \label{Fig:HiFI_zoom}
\end{figure}

\subsection{Imaging spectroscopy in the visible}
Imaging spectroscopy along the \ion{Na}{i} D$_2$ 5889.95\,\AA\ line (see Table \ref{tab:filters}) was 
carried out with the GREGOR Fabry-P\'erot Interferometer \citep[GFPI,][and references therein]{puschmann12}.
The \ion{Na}{i} D$_2$ line maps the upper photosphere to lower chromosphere in the solar atmosphere. 
The scan along the spectral line comprised 16 wavelength positions on an equidistant grid with a 
spectral sampling of 56.5\,m\AA. The number of images per wavelength position was eight and the
exposure time was 40\,ms. 
Between 08:38--09:07\,UT 52 scans, with a cadence of 31.6\,s, were acquired. The GFPI setup included
two synchronized CCD cameras, which were located in the narrow- and broad-band channels of the GFPI. 
This configuration allowed for image restoration using 
multi-object multi-frame blind deconvolution \citep[MOMFBD,][]{lofdahl02,vannoort05}. 
Before restoration, all images were corrected by the average dark image and divided by the 
flat field. The FOV after image restoration was $49\farcs1 \times  35\farcs8$, with a spatial 
sampling of 0.04\arcsec\,pixel$^{-1}$ and no binning. The MOMFBD-restored narrow-band images were finally corrected for the 
prefilter transmission curve and the blueshift across the FOV. 
The data reduction and image restoration procedures are part of AIP's software 
package \texttt{sTools}\footnote{http://gregor.aip.de/ (requires free registration)} \citep{stools}.
The blue wing of the \ion{Na}{i} D$_2$ line at $\lambda_0-0.22$\,\AA\ reflects well the location of the
BPs (lower left panel of Fig. \ref{Fig:overview}).

\subsection{Very fast imaging in the violet and blue}
The highest spatial resolution was achieved in the violet and blue spectral range using the 
two synchronized sCMOS cameras of the High-resolution Fast Imager \citep[HiFI,][]{stools,denker18b}. 
Table \ref{tab:filters} describes the properties of the two interference filters. 
The spatial sampling was 0\farcs025\,pixel$^{-1}$. Nevertheless, the angular resolution $R$  
of the telescope, following equation $R = \lambda/D$ (where $\lambda$ is the wavelength and
$D$ is the effective diameter of the primary mirror), yields
$\sim$0\farcs057 and $\sim$0\farcs065, for \ion{Ca}{ii} H and blue continuum, 
respectively. Hence, we oversampled by a factor of about 1.14 and 1.30, respectively.

\begin{table}[!ht]
\begin{center}
\caption{Interference filters for HiFI and GFPI.}\label{tab:filters}
\begin{tabular}{ccccc}
\hline
\hline
Instrument             & Line                  & Filter (\AA) & FWHM (\AA) \rule[-4pt]{0pt}{15pt}  \\
\hline
 \multirow{2}{*}{HiFI} & Blue continuum        & 4505             & 12.0  \rule[-4pt]{0pt}{14pt}  \\
                       & \ion{Ca}{ii} H        & 3968             & 11.0    \\
\hline
GFPI                   & \ion{Na}{i} D$_2$     & 5892             & 6.7  \rule[-4pt]{0pt}{14pt} \\
\hline
\end{tabular}
\end{center}
\end{table}

The observing strategy with HiFI consisted in repeated bursts of 500 images 
with a frame rate of 47\,Hz and an exposure time of
7\,ms (between 08:38--09:00\,UT) and 6\,ms (between 09:00--09:07\,UT), 
to adjust for intensity changes because of the rising Sun. A series of 
500 images was acquired within 23\,s. 
Dark and flat-field corrections were carried out as part of the data reduction pipeline 
\texttt{sTools}. 
Afterward, the 100 best images of each series  
were identified using the Median Filter-Gradient Similarity (MFGS) algorithm \citep{mfgs,denker18a}.
The selected images were then restored using the speckle-interferometry code
KISIP \citep{kisip}.

The upper panels of Fig. \ref{Fig:overview} exhibit speckle-restored HiFI images. Bright points are
straightforwardly identified in the \ion{Ca}{ii} H filtergram, whereas the blue continuum image shows less contrast
between the BP and its surroundings. The regions of interest and a comparison between 
the isolated BPs under study in both filters is depicted in Fig. \ref{Fig:HiFI_zoom}.

\section{Data analysis} \label{Sect:3}

\subsection{Alignment}
The spatial resolution and FOV of the three instruments is different.
In addition, difficulties arise when 
matching slit-reconstructed images, like the ones acquired with GRIS, to 
imaging instruments such as GFPI and HiFI. The spatial sampling of GRIS is 
about 5.4 times larger than the one of HiFI. To accomplish this task, calibration 
images of the United States Air Force (USAF) resolution target inserted at GREGOR's focus F3 were used. 
The standard rotation and correlation procedures from 
the Interactive Data Language (\texttt{IDL}), 
as well as the alignment routine for proper motion measurements from \citet{molowny94}
and \citet{yi95}, which belongs to the \texttt{oslolib}, were used 
to successfully match the images from different instruments. Furthermore,
a visual inspection was carried out to verify their coincidence, 
since all analyzed BPs appear as enhancements of the intensity for all wavelengths.

\subsection{Inversions}
The \ion{Si}{i} 10827\,\AA\ line was inverted using the 
Stokes Inversions based on Response functions \citep[SIR;][]{SIR} code.
The abundance was taken from \citet{grevesse84} (7.50 for Si) and the
logarithm of the oscillator strength times the multiplicity of the lower level, 
$\log g f = 0.363$, from \citet{borrero03}. 
The used collisional broadening parameters $\alpha$ and $\sigma$, which result
from the quantum mechanical theory of Anstee, Barklem, and O'Mara, were
0.231 and 2.0412$\times$\,$10^{-14}$, respectively.
The SIR code assumes local thermodynamical equilibrium (LTE) and hydrostatic
equilibrium to iteratively solve the radiative transfer equation. 
The code provides height-dependent physical parameters. The height stratification 
is expressed in logarithmic units of the optical depth ($\log \tau$) at 5000\,\AA. Up to 
four nodes across the atmospheric height were used. The microturbulence was kept constant with height. 
A single magnetic component was used together with a non-magnetic component (only intensity). 
The latter serves in the inversions to infer the filling factor $f$.
The filling factor accounts for the fraction of magnetic component within the resolution 
element of the observations and is used to determine the magnetic field strength.
The use of a filling factor is necessary for the \ion{Si}{i} data, as we are not fully
resolving the BPs. This can be clearly seen by comparing the high-resolution \ion{Ca}{ii} H and 
blue-continuum filtergrams to the slit-reconstructed \ion{Si}{i} map. 
The non-magnetic component was computed by averaging the Stokes $I$ profiles of a quiet-Sun area within the FOV.
Two magnetic components were used for a few pixels to disentangle the atmosphere that produced
some specific asymmetries in the Stokes $V$ profiles near BPs (see Sect. \ref{Sect:asymStokesV}).

The inversions were initialized with three different model atmospheres: 
the FAL-C and FAL-F models \citep{fontenla06} and the Harvard-Smithsonian reference 
atmosphere model \citep[HSRA;][]{gingerich71}. The FAL-C and HSRA models represent 
quiet regions of the Sun, whereas FAL-F simulates an active network. 
The models cover an optical depth range at 5000\,\AA\
between $1.4 \geq \log \tau \geq -4.0$. The resulting atmospheres, which best fit 
the observed Stokes profiles were stored. The evaluation of the fits was performed by
using a $\chi^2$-test, which computes the sum of the squared differences between the observed and 
synthesized Stokes profiles.

Particular attention needs to be drawn to the line core of the \ion{Si}{i} line,
since it is formed under non-LTE (NLTE) conditions. The main
NLTE effect is that the line core is deeper than under LTE conditions
\citep{bard08}. 
Response functions to a given quantity -- magnetic field and LOS 
velocity -- allow to determine at which optical depths range 
the spectral line is sensitive to perturbations on such quantity.
The response
functions were calculated using one atmospheric model obtained from the inversions of each bright point 
(a--c). It is found that the \ion{Si}{i} line is sensitive to changes in the atmosphere
below $\log \tau \lesssim -1.0$ until almost $\log \tau \sim -4.0$. 
\citet{shchukina17} found NLTE effects also in Stokes $Q$, $U$, and $V$ in the inner wings of the \ion{Si}{i} line,
around $\Delta \lambda = 0.1$\,\AA. A comparative study inverting the \ion{Si}{i} line with and without
departure coefficients from LTE with SIR demonstrated that rms variations of the field strength (up to 150\,G) and
the LOS velocity (up to 0.2\,km\,s$^{-1}$) exist at $\log \tau = -0.2$ \citep{kuckein12a}. 
We consider these variations rather low and will constrain our analysis between $-1.0 \geq \log \tau \geq -2.5$, 
hence avoiding smaller optical depths. 

\begin{figure}[!t]
 \centering
 \includegraphics[width=\hsize]{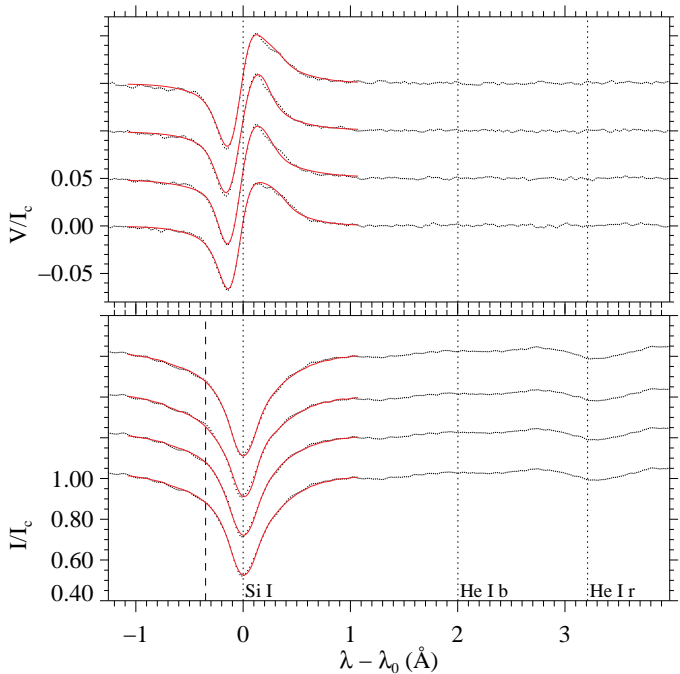}
\caption{Four Stokes $V/I_\mathrm{c}$ profiles with the largest amplitude belonging to BP (b), and their 
  corresponding Stokes $I/I_\mathrm{c}$ profiles. The various Stokes $I$ ($V$) profiles are shifted by 
  $+0.20$ ($+0.05$) along the $y$-axis or ordinate. The wavelength axis is centered at the \ion{Si}{i}
  line at rest. The red line shows the best fit obtained from the SIR inversion. The dotted vertical 
  lines show the positions at rest of \ion{Si}{i} and of the blue (\ion{He}{i} b) and red (\ion{He}{i} r)
  components of \ion{He}{i}. The dashed vertical line indicates \ion{Si}{i} 
  $\lambda_0-0.35$\,\AA, which corresponds to the image shown in the lower right corner of Fig. \ref{Fig:overview}. 
  }
  \label{Fig:stokes_BPs}
\end{figure}

\section{Results} \label{Sect:4}

\subsection{Bright point fine structure}

The blue continuum at 4505\,\AA\ and \ion{Ca}{ii} H images from HiFI reveal small-scale details in and around 
the BPs. Rather than point-like features, the observed BPs resemble chains of BPs which appear as 
elongated tube-like structures (Fig. \ref{Fig:HiFI_zoom}). The length of individual BPs 
lies in the subarcsecond range, whereas chains of BPs extend a few seconds of arc. Moreover, the thickness of isolated 
BPs, as shown in Figs. \ref{Fig:HiFI_zoom}(a) and \ref{Fig:HiFI_zoom}(c),
corresponds to 0\farcs30--0\farcs35 (217--253\,km). They lie along intergranular lanes and are very
dynamic. Time-evolution movies (not shown) exhibit that their shape is molded by the expansion and 
displacement of the adjacent granules, which are much larger than the BPs themselves. A detailed analysis
of the filtergram time evolution is beyond the scope of this work and is postponed to a future paper. 

\begin{figure*}[!t] 
   \resizebox{\hsize}{!}{\includegraphics{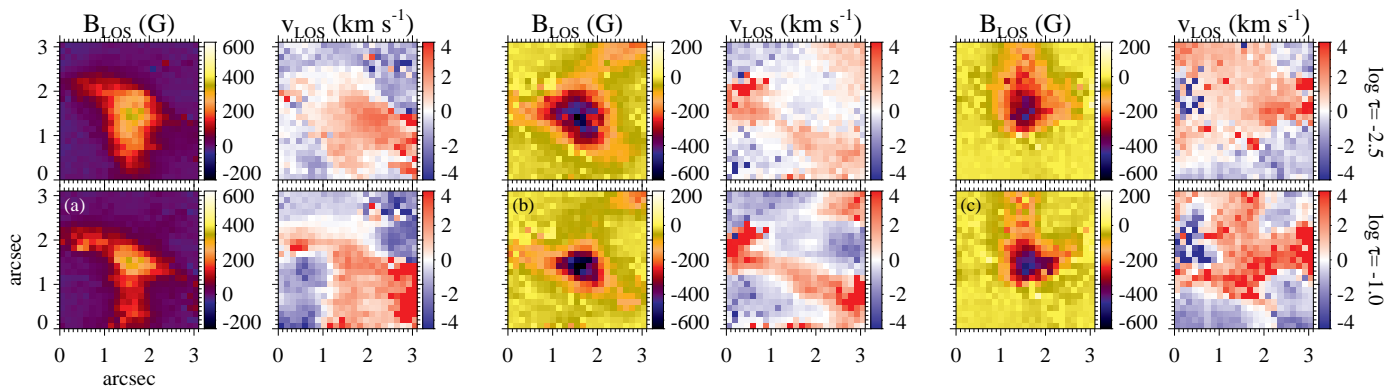}}
      \caption{Longitudinal magnetic field and velocity maps inferred 
      from the \ion{Si}{i} 10827\,\AA\ inversions of the three 
      isolated BPs at two different optical depths, \mbox{$\log \tau = -1.0$} (bottom) and $-2.5$ (top). 
      Bright points (a)--(c) are shown from left to right. We note that BP (a) has opposite polarity
      as compared to BPs (b) and (c). Magnetic fields are clipped between $-$200\,G and $+$600\,G for BP (a), and 
      between $-$600\,G and $+$200\,G for BPs (b--c). LOS velocities are clipped between $\pm 4$\,km\,s$^{-1}$. Negative
      velocities represent flows along the LOS toward the observer, that is, upflows 
      because of the proximity to disk center. The FOV corresponds to a zoom of the region-of-interest
      (white box in Fig. \ref{Fig:HiFI_zoom}).
      }
\label{Fig:inversion_maps}
\end{figure*}

\subsection{Bright point infrared Stokes profiles}
Bright points were easiest identified in the high-spatial resolution HiFI broad-band blue-continuum 
and \ion{Ca}{ii} H images. Afterward, the corresponding area was scrutinized in the near-infrared data. 
Interestingly, BPs are also well distinguished when selecting the blue wing of the \ion{Si}{i} line in a
slit-reconstructed image, at about $\lambda_0-0.35$\,\AA\ away from the line core. 
The fact that BPs appear bright in the wing of the \ion{Si}{i} line, and not in the line core, indicates that 
the perturbation is located in deep layers of the photosphere, because the \ion{Si}{i} line core is formed 
in the upper photosphere.
The same applies to the \ion{Na}{i} D$_2$ line, where BPs are best detected in filtergrams of the blue wing, 
at about $\lambda_0-0.22$\,\AA\ away from the line core. 

In the three isolated BPs, the Stokes $V$ profiles with the largest amplitude match the brightest intensity in 
\ion{Si}{i} $\lambda_0 - 0.35$\,\AA\ (e.g., Fig. \ref{Fig:twocompmaps}). 
An example showing the four Stokes profiles with highest signal (up to $V/I_\mathrm{c}$$\sim$0.07) 
in BP (b) is presented in 
\mbox{Fig. \ref{Fig:stokes_BPs}}. In this figure, the spectral profiles were slightly smoothed by 
convolving them with a normalized
Gaussian with a FWHM of $\sim$36\,m\AA, which roughly corresponds to the spectral resolution. 
The red line shows the best fit from the SIR inversions. The dashed vertical line in the lower panel marks the
spectral position of \ion{Si}{i} $\lambda_0 - 0.35$\,\AA, where BPs are most easily identified. The dotted vertical lines
mark the wavelengths at rest of \ion{Si}{i} 10827\,\AA\ and the blue (\ion{He}{i} b) and blended red (\ion{He}{i} r) 
components of the \ion{He}{i} 10830\,\AA\ triplet. 

Upon closer inspection, the \ion{Si}{i} profiles of BP (b) shown in Fig. \ref{Fig:stokes_BPs} exhibit asymmetries.
The Stokes $I/I_\mathrm{c}$ red wing
appears more tilted compared to the blue wing and Stokes $V/I_\mathrm{c}$ shows a much broader and extended red
lobe compared to the blue one. A likely explanation for this is the existence of gradients in the 
physical parameters along the LOS, that is, the quantities vary across different optical depths of the solar atmosphere. 
Stokes $Q$ and $U$ were not taken into account, as their profiles were at the noise level and therefore 
not discernible.
The standard deviation of the Stokes $V/I_\mathrm{c}$, $Q/I_\mathrm{c}$, and $U/I_\mathrm{c}$ profiles 
computed in the continuum was around $10^{-3}$, which is of the same order of magnitude as the standard deviation of the 
Stokes $Q/I_\mathrm{c}$ and $U/I_\mathrm{c}$ profiles in the spectral line itself. 

No polarization signatures were detected in the \ion{He}{i} profiles of the selected BPs (Fig. \ref{Fig:stokes_BPs}). 
Stokes $V/I_\mathrm{c}$ appears completely flat 
and the red component of Stokes $I/I_\mathrm{c}$ shows some, but very low, absorption. Hence, 
if polarization signals were present, they were very weak, below the noise level of our data.

\subsection{Magnetic field structure and Doppler velocities}
Spectral line inversions of the \ion{Si}{i} Stokes profiles were carried out for the three selected isolated BPs. 
We concentrated on a zoom of the region-of-interest in GRIS of about $3\arcsec \times 3$\arcsec, which generously covered 
the small BPs (white box in Fig. \ref{Fig:HiFI_zoom}). 
Solely Stokes $I$ and $V$ were taken into account, as the Stokes signals of $Q$ and $U$ were at the noise level.
Furthermore, the observed FOV was close to disk center, which assures that the differences 
between the LOS magnetic field and the vertical magnetic 
field in the solar reference frame ($B_\mathrm{LOS} \approx B_\mathrm{vertical}$) are small. 
Vertical magnetic fields refer to radially oriented fields.

The longitudinal magnetic field in each pixel was computed using the equation
\begin{equation} \label{Eq:1}
 B_\mathrm{LOS} = f |B| \cos \gamma \hspace{0.1cm},
\end{equation}
taken from \citet{landi92}, where $|B|$ is the magnetic field strength, $f$ the filling factor, and $\gamma$ the  
LOS inclination of the magnetic field. All three parameters were inferred during the SIR inversions. 
The inferred filling factor in the area of the BPs was, on average, in the range of 0.3--0.5. 

The longitudinal magnetic field at two different optical depths inferred from the \ion{Si}{i} line is 
shown in Fig. \ref{Fig:inversion_maps}.
From left to right, the isolated BPs (a)--(c) are depicted. The lower row corresponds to 
a larger optical depth in the solar atmosphere ($\log \tau = -1.0$), whereas the top row corresponds to 
$\log \tau = -2.5$. The first column shows BP (a), which has an inverted color table with respect to
BPs (b)--(c), because of the positive polarity. A visual inspection comparing $B_\mathrm{LOS}$ 
between the two optical depths reveals that the field covers a larger fraction of the FOV at $\log \tau = -2.5$.
On the contrary, at $\log \tau = -1.0$, the field is more concentrated. 
Let us further quantify this expansion. Therefore, we identified pixels within the FOV of Fig. 
\ref{Fig:inversion_maps} close to the BPs, which had $B_\mathrm{LOS} > 100$\,G or $B_\mathrm{LOS} < -100$\,G,
depending on the polarity of the BP. This value is a good threshold to encompass the LOS magnetic field
for all BPs. We found an expansion of the magnetic field 
with height in all three BPs. In particular, the smallest extension was found for 
BP (a), with an increment of about 7\% of the pixels fulfilling $|B_\mathrm{LOS}| > 100$\,G.
We note that BP (a) has also the smallest size of all the studied BPs (Fig. \ref{Fig:HiFI_zoom}). 
The largest increment of the area of the
magnetic field by 51\% was seen in BP (b), followed by 19\% for BP~(c). 

\begin{figure}[!t]
 \centering
 \includegraphics[width=\hsize]{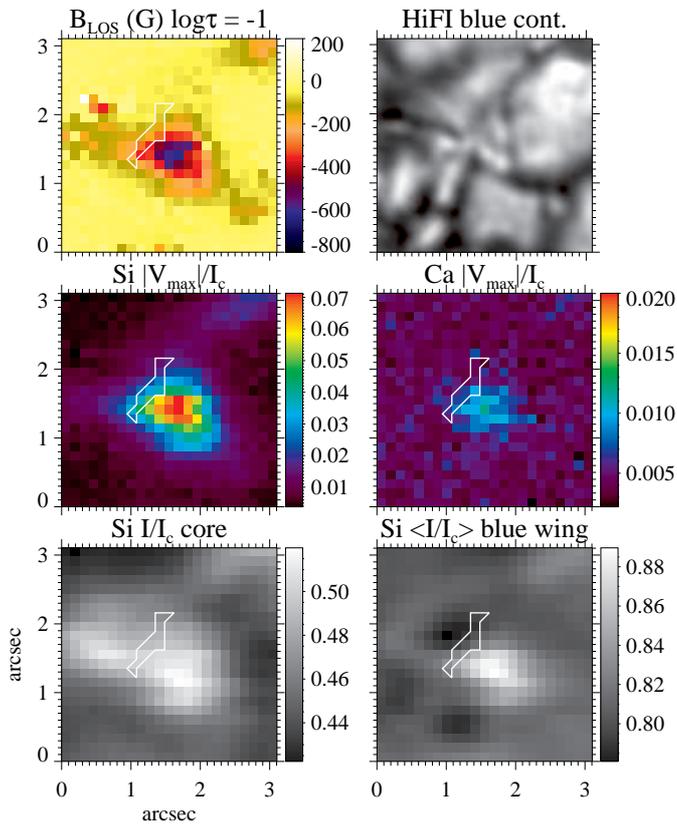}
\caption{Zoom of the region-of-interest of $3\arcsec \times 3\arcsec$ belonging to BP (b) from Fig. \ref{Fig:HiFI_zoom}. 
  The lower two panels show the intensity of the \ion{Si}{i} line in the 
  core (left) and averaged blue wing at around $\lambda_0-0.35$\,\AA\ (right), where BPs are well detected. 
  The middle left (right) panel depicts the unsigned maximum signal in Stokes $V_\mathrm{max}/I_\mathrm{c}$
  from the \ion{Si}{i} 10827\,\AA\ (\ion{Ca}{i} 10839\,\AA) line.
  The upper left panel represents the longitudinal magnetic field at $\log \tau = -1$ inferred from the \ion{Si}{i}
  inversions. A zoom of the HiFI blue-continuum image showing the fine structure is exhibited in the 
  upper right panel.
  Pixels enclosed inside the white contours show atypical 
  \ion{Si}{i} Stokes $V$ profiles, with two discernible red lobes like the example shown 
  in \mbox{Fig. \ref{Fig:twocompprofiles}}.   
  }
  \label{Fig:twocompmaps}
\end{figure}

We repeated a similar experiment comparing the circular polarization signal of the
photospheric \ion{Ca}{i} 10839\,\AA\ line to the \ion{Si}{i} line. 
The former is formed in a rather thin layer corresponding to the deep 
photosphere \citep[e.g.,][]{felipe16}, 
while the latter in the upper photosphere. 
The middle panels of Fig. \ref{Fig:twocompmaps} 
compare the absolute maximum amplitude of Stokes $V_\mathrm{max}/I_\mathrm{c}$ among both lines.
The amplitude of the circular polarization arising from the \ion{Ca}{i} line is much smaller
than the \ion{Si}{i} one. However, this is mainly due to the fact that the spectral line is 
much shallower than the deep \ion{Si}{i} line. What is more important is the spatial 
extension of the Stokes $V_\mathrm{max}/I_\mathrm{c}$ signals in both lines. 
The \ion{Ca}{i} Stokes $V_\mathrm{max}/I_\mathrm{c}$ signals appear compact around the BP,
whereas the \ion{Si}{i} counterpart shows a widespread area centered at the BP.
This wider area of the \ion{Si}{i} signal provides further evidence of the expansion 
of the magnetic field lines with height at the isolated BP.  

The order of magnitude of the longitudinal magnetic fields inferred from the \ion{Si}{i} inversions 
($B_\mathrm{LOS}$) is computed for the three isolated BPs. For this purpose, 
an average of $5 \times 5$ pixels (about 0\farcs7$\times$0\farcs7) centered at the BPs is calculated 
and then averaged over both heights 
($\log \tau = -1.0$ and $-2.5$). The mean LOS magnetic field is in the range of 220--340\,G. 
Individual pixels, with magnetic fields along the LOS of up to 620\,G, were found in BP~(b).  

The \ion{Si}{i} inferred LOS velocity maps reveal that the studied isolated BPs are associated with downflows (redshifts). 
This is not surprising as BPs are located in intergranular lanes, between granules which have 
rising plasma in their center. The average LOS velocities centered in the same abovementioned box of $5\times5$ pixels 
for the three BPs is in the range of 0.6--3.1\,km\,s$^{-1}$ for $\log \tau = -1.0$. 
Moreover, for $\log \tau = -2.5$ we find average LOS velocities between 0.0--1.8\,km\,s$^{-1}$. 
In BPs (b) and (c), the mean velocities 
clearly decrease with height, whereas BP (a) shows no variation with height, also not in its standard 
deviation within the $5\times5$ pixels box. 
Nevertheless, these results have to be interpreted with caution as the Stokes profiles in the BPs
show asymmetries which are likely produced by different flows along 
the LOS (Fig.~\ref{Fig:stokes_BPs}). We will investigate this in the next section.

\subsection{Highly redshifted Stokes V lobes} \label{Sect:asymStokesV}

\begin{figure}[!t]
 \centering
 \includegraphics[width=\hsize]{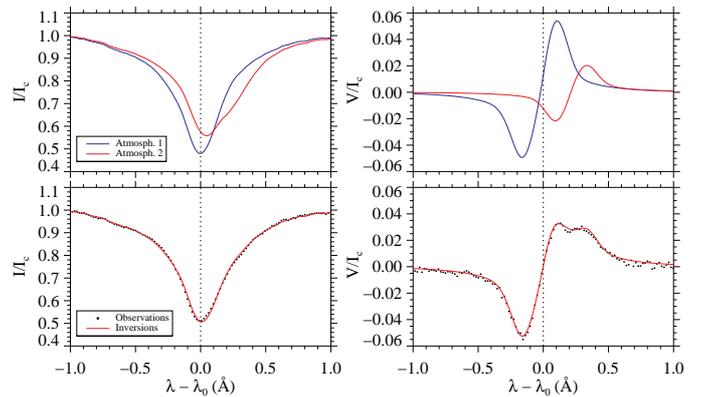}
\caption{Representative example of Stokes $I/I_\mathrm{c}$ and $V/I_\mathrm{c}$ profiles 
  of a highly redshifted pixel close to the center of BP (b). The profiles are taken from inside of 
  the white contour of Fig. \ref{Fig:twocompmaps}. 
  The lower two panels show the observations (black dots) and best fit (solid red line)
  from a two-component SIR inversion. The upper two panels depict the synthesized profiles arising from
  the two atmospheric models obtained with SIR which coexist within the same resolution element. 
  Atmosphere 1 (blue solid line) and 2 (red solid line) have a filling factor of about 0.7 and 0.3, 
  respectively. The inferred average LOS velocity and inclination between $-1.0 \geq \log \tau \geq -2.5$ 
  from this particular fit is $v_\mathrm{LOS,1} \sim -1$\,km\,s$^{-1}$ and $\gamma_1 \sim 172^\circ$ and 
  $v_\mathrm{LOS,2} \sim 6$\,km\,s$^{-1}$ and $\gamma_2 \sim 169^\circ$, for atmospheres 1 and 2, respectively.
  }
  \label{Fig:twocompprofiles}
\end{figure}

All three isolated BPs reveal signs of asymmetric Stokes $V$ profiles. 
This becomes evident not only when inspecting the profiles in the center of the BP (Fig. \ref{Fig:stokes_BPs}), 
but also even clearer a few pixels away from the center (Fig. \ref{Fig:twocompmaps}). For example, for BP (b), 
two distinguishable red lobes in the Stokes $V$ profile are found within the white contour in Fig. \ref{Fig:twocompmaps}.
An example of the shape is
depicted in Fig.~\ref{Fig:twocompprofiles}. The observed intensity profile in the lower left
panel is not symmetric, having a red wing which is wider than the blue wing. 
Inspecting the Stokes $V$ profile reveals that another magnetized component is 
inside the resolution element. In order to disentangle the two signals for these pixels, 
we carried out inversions with two magnetized components and one non-magnetic component, 
to deduce the velocity of the 
highly shifted component. The inversions were initialized with a slightly different setup as the regular
single magnetic component inversions. More weight was given to Stokes $V$ to assure 
a good fit to the second magnetic component. To decrease the number of free parameters and focus on the
Doppler shift of the second component, the macroturbulence was fixed in both model atmospheres 
to the theoretical value of the spectral resolution of the data ($\sim$1 km\,s$^{-1}$), and only two nodes
were used for magnetic field strength, LOS velocity, and inclination of the field. 
The filling factor was left as a free parameter and served as a parameter to weight the importance of each 
magnetized atmosphere inside of the pixel. The outcome of the inversion is presented 
in Fig. \ref{Fig:twocompprofiles} as a red solid line in the lower two panels. The Stokes 
$I/I_\mathrm{c}$ and $V/I_\mathrm{c}$ profiles which arise from each individual atmosphere, 
weighted by their corresponding
filling factor in the case of Stokes $V/I_\mathrm{c}$, are displayed in the upper panels. 
The coexistence of two magnetic components is clearly evident within the same resolution element. 
Both components share similar magnetic field and inclination values, 
but significantly differ in their Doppler shifts. 
When averaging the LOS velocities across 
$-1.0 \geq \log \tau \geq -2.5$ (16 layers in optical depth), a blue shifted component with
a LOS velocity of $v_\mathrm{LOS,1} \approx -1$\,km\,s$^{-1}$ and a standard deviation (across the
16 layers of optical depth) of 
$\sigma_1 = 0.16$\,km\,s$^{-1}$ is found. Conversely, a strongly redshifted component with a  
LOS velocity of $v_\mathrm{LOS,2} \approx 6$\,km\,s$^{-1}$ and $\sigma_2 = 0.08$\,km\,s$^{-1}$
is retrieved (red solid line in the upper panels of Fig. \ref{Fig:twocompprofiles}). 
The blueshifted magnetic component has a larger filling factor ($f_1$$\approx$0.7) compared to the
highly redshifted one ($f_2$$\approx$0.3). Hence, the later contributes less to the observed
Stokes profiles.

\begin{figure*}[!t]
\sidecaption
  \includegraphics[width=12cm]{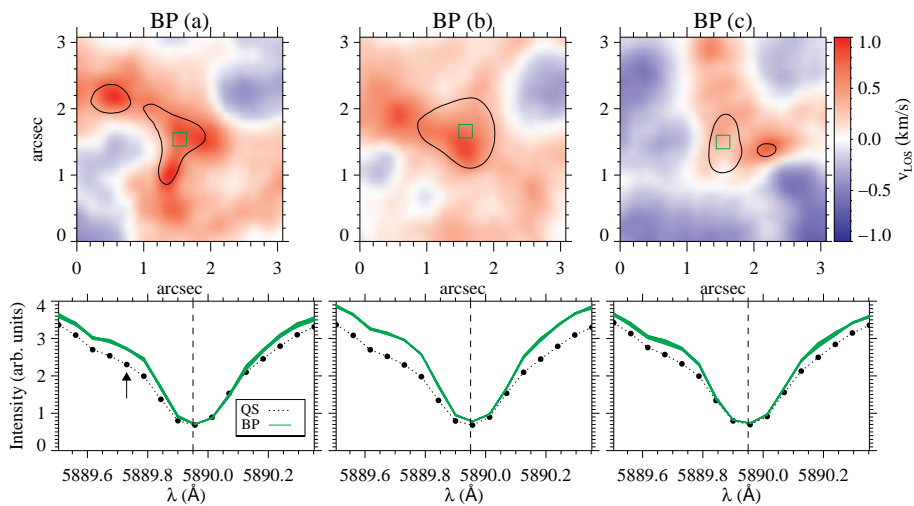}
     \caption{Upper row: Doppler shifts for BPs (a)--(c) inside the zoomed region-of-interest,
     using the bisector method 
     at a line depth, which corresponds to about \ion{Na}{i} D$_2$ $\lambda_0 - 0.22$\,\AA. 
     The arrow in the lower left panel illustrates the depth at which the bisector was 
     computed. Black contours mark the area of the BP. 
     The lower panels show the average quiet-Sun profile (dotted line) within 
     the whole FOV (Fig. \ref{Fig:overview}). The large dots correspond to the acquired
     filtergrams along the \ion{Na}{i} D$_2$ line with GFPI. The green solid lines exhibit the brightest
     25 profiles belonging to the green box in the upper panels. 
     The dashed vertical line marks the wavelength at rest.
     }
     \label{Fig:NaD2velocities}
\end{figure*} 
 
\subsection{Na {\small I} Doppler shifts} \label{Sect:NaD2velocities}

Simultaneous spectral information of the \ion{Na}{i} D$_2$ line in the abovementioned BPs is provided by 
the GFPI. The closest scan in time for each BP was used to compute the Doppler shifts at a line 
depth which corresponds to $\lambda_0 \pm 0.22$\,\AA. 
The reason for choosing this particular filtergram is because BPs show their highest contrast at 
$\lambda_0 - 0.22$\,\AA\ when scanning with the GFPI along the \ion{Na}{i} D$_2$ line. The bisector method 
was used to infer the shift of the inner wings of the line with respect to the quiet-Sun reference profile, 
which in turn was determined by averaging areas within the whole FOV, which were absent 
of dark and bright structures. Line-core fits or fits of the whole spectral line with a Gaussian
will yield different velocities. The former will map higher layers in the atmosphere, while the latter
will provide an average shift of the whole spectral line. 

The upper panels of Fig. \ref{Fig:NaD2velocities} illustrate the LOS velocities in a small FOV 
of about $3\arcsec \times 3\arcsec$ centered at each BP. 
The depth of the line at which the bisectors
were computed is marked with a black arrow in the lower left panel. It corresponds to the fifth
wavelength position when scanning along the line  
(roughly at $\lambda_0 - 0.22$\,\AA\ of the \ion{Na}{i} D$_2$ line). 
The black contours mark intensity enhancements associated with the BPs. The three isolated BPs show a common 
velocity pattern, which is dominated by moderate-to-low downflows. 
BPs (a) and (b) exhibit the largest velocities, sporadically reaching up to 1\,km\,s$^{-1}$.  

A closer view of the spectral profiles at the BPs is provided in the lower three panels of 
Fig. \ref{Fig:NaD2velocities}. The dotted curve represents the quiet-Sun profile. 
The \ion{Na}{i} D$_2$ intensity profiles belonging to the green box in the upper 
LOS velocity maps are shown in solid green. The green box, centered at the 
highest intensity of the BP, comprises 25 pixels. The highest intensity refers to the highest value at the wavelength
position corresponding to the black arrow in Fig. \ref{Fig:NaD2velocities}. One interesting
observational feature is the enhancement of the inner blue wing. This naturally explains
why BPs are best seen in filtergrams corresponding to $\lambda_0 - 0.22$\,\AA.
The enhancement in the blue inner wing appears in all three isolated BPs. 
Conversely, the line core intensity remains close to the one of the quiet-Sun reference profile.

\begin{figure*}[!t]
  \resizebox{\hsize}{!}{\includegraphics{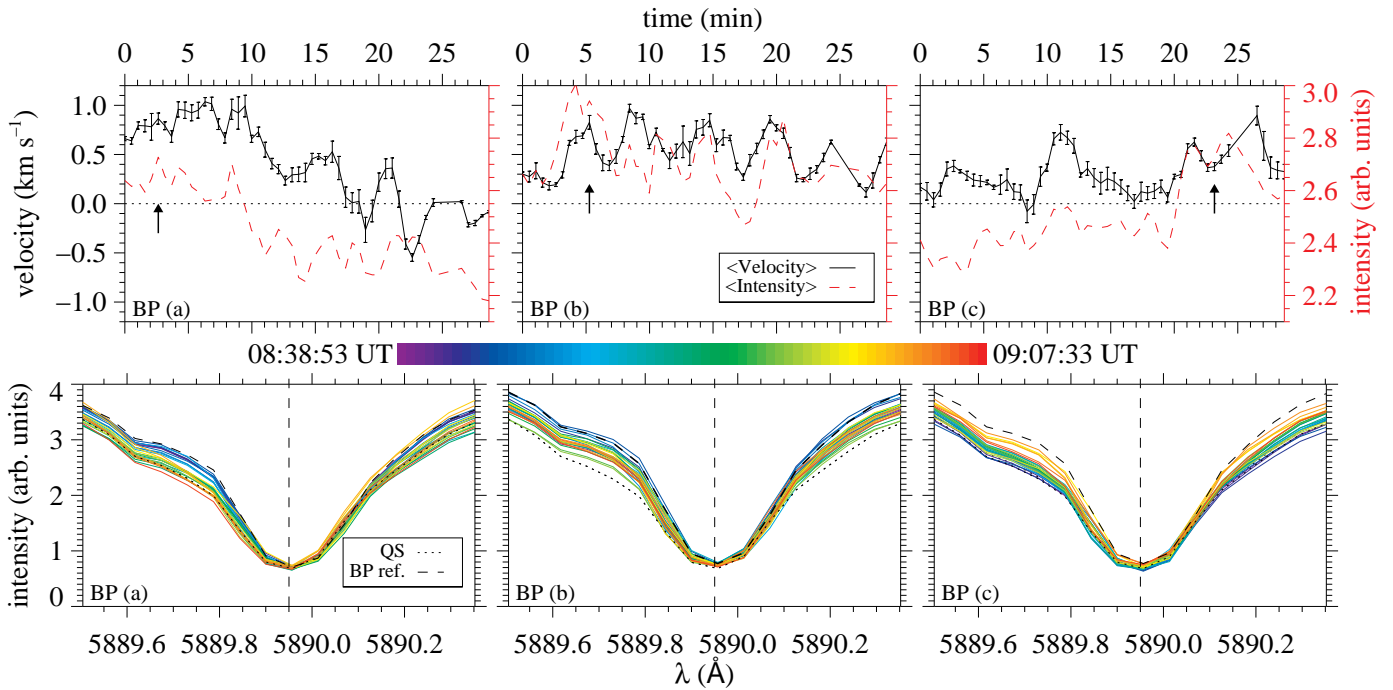}}
     \caption{Upper row: time evolution of the average LOS velocities computed within the 
     $5 \times 5$\,px$^2$ green box shown in Fig. \ref{Fig:NaD2velocities}. The vertical bars indicate $\pm$ the
     standard deviation of the velocities within the same box. The overplotted red dashed line shows the average intensity
     in the blue wing of \mbox{\ion{Na}{i} D$_2$} inside of the same green box. The arrow marks the time of simultaneous GRIS data. 
     Lower row: average intensity profile generated within the same green box. The color code indicates 
     the time of each profile. The dashed vertical line marks the wavelength at rest. The dotted profile 
     represents the quiet-Sun profile whereas the dashed one corresponds in time to the average profile marked
     with the arrows in the above panels. }
     \label{Fig:Nad2timeevol}
\end{figure*}

The temporal evolution of the average spectral profile within the green box in Fig. \ref{Fig:NaD2velocities} is
depicted for each BP in the lower panels of Fig. \ref{Fig:Nad2timeevol}. The color table maps the different times from the
beginning of the observations (violet) toward the end of the time series (red). 
As a comparison, the quiet-Sun profile is represented with a dotted line. 
In addition, the dashed profile matches in time with the \ion{Si}{i} observations. 
The line core does not show significant changes in intensity during the whole times series. Conversely, 
the wings of the line
appear vertically shifted in intensity, especially in the blue wing. This is the reason why it was used as 
an indicator for BPs. Overall the profiles do not manifest large Doppler shifts. 

The upper panels of Fig. \ref{Fig:Nad2timeevol} track the average LOS velocity (black line) and intensity 
(red dashed line) of the BPs within the same
green box marked in Fig. \ref{Fig:NaD2velocities}. While the solid line depicts the average velocity, the bars
indicate one $\sigma$ (standard deviation within the same box) up and down. 
This information reveals how much the velocities fluctuate in the BP. Bright point (a) shows the best correlation
between the LOS velocity and the intensity. Both follow a similar decreasing trend. 
The velocities are larger when the BP is prominently seen (higher intensity).
The intensity and LOS velocity in BP (b) oscillates, but is overall persistent. As opposed to BP (a), BP (c) seems to
be slowly brightening up, as indicated by the positive trend of the intensity. Often, but not always, peaks in intensity 
are correlated with peaks in velocity. For instance, the intensity peak at around 11 and 21\,min has a redshift
of about 0.7\,km\,s$^{-1}$ and 0.6\,km\,s$^{-1}$ on average, respectively.
The standard deviation of the LOS velocities inside the region of interest is generally small. It only oscillates 
between 0.01--0.16\,km\,s$^{-1}$ during the almost 30 minutes of observations.

\section{Discussion} \label{Sect:5}

\subsection{Identification of bright points}

The current multi-wavelength study shows that BPs are identified in a variety 
of wavelengths. From our simultaneous observations, BPs are best recognized in the broad-band 
(FWHM$=$11\,\AA) \ion{Ca}{ii} H filtergrams, which is the highest 
angular-resolution data set of the present work. 
In addition, their fine structure appears 
in the blue continuum filtergrams, though with less contrast. 
Caution has to be taken, since small elongated structures, similar to BPs, also
appear in blue continuum images but with neither a \ion{Ca}{ii}~H counterpart
nor polarization signals in the near-infrared lines. 
A good example is seen in Fig. \ref{Fig:HiFI_zoom}, 
at the middle panel at coordinates ($x,y$)$=$(3\farcs5, 3\farcs4). This pseudo-BP is almost touching
the BP in the blue-continuum filtergram, whereas it is not detectable in the
\ion{Ca}{ii}~H filtergram. The question remains if these pseudo-BPs will appear
bright in G-band filtergrams, which are typically used to pinpoint BPs.

Furthermore, BPs are well identified in the blue wings of 
\ion{Si}{i} at $\lambda_0-0.35$\,\AA\ and \ion{Na}{i} D$_2$ at $\lambda_0-0.22$\,\AA\
(Fig. \ref{Fig:overview}). This seems to be a common property of other 
photospheric lines such as the \ion{Fe}{i} 5250.5\,\AA\ line \citep{utz14}.
Yet, deeply formed lines such as the \ion{Ca}{i} 10839\,\AA\ line, also trace BPs
in line-core images. 
On the contrary, higher up in the atmosphere, the chromospheric 
\ion{He}{i} 10830\,\AA\ triplet remains unaffected, at least in our studied isolated BPs. 
In summary, the present data provide evidence of the low-atmospheric
nature of BPs. Although BPs were identified in the line core of \ion{Na}{i}~D$_1$ (which has a
similar behavior to the \ion{Na}{i}~D$_2$ line analyzed in this work), the core rather 
samples magnetic concentrations in the photosphere, not in the chromosphere \citep{leenaarts10}.
\citet{jess10} reported on BPs in the line core of \ion{Na}{i}~D$_1$. Their spectral range 
was just half of the range observed in the present \ion{Na}{i}~D$_2$ line, 
and the deduced Doppler shifts reached up to 7\,km\,s$^{-1}$.
We conjecture that their intensity enhancements were mainly related to the strong redshifts of the
whole \ion{Na}{i}~D$_1$ line profile, which then showed the inner blue wing. A LOS velocity 
of 7\,km\,s$^{-1}$ corresponds to a Doppler shift of about $0.14$\,\AA\ for this line.

\subsection{Doppler velocities in the bright points}
Height-dependent inversions of the \ion{Si}{i} 10827\,\AA\ line revealed downflows 
at the three isolated BPs. The LOS velocities were smaller 
at lower optical depths of the atmosphere, at $\log \tau = -2.5$, in two BPs, whereas the other BP
remained constant with height. However, when scrutinizing 
the individual Stokes profiles, we noticed asymmetries in Stokes $I$ and $V$.
Such asymmetries are not uncommon in BPs  \citep[e.g.,][]{riethmuller14,martinezgonzalez2012}.
In our case, in some of the Stokes $V$ profiles, two components became clearly 
discernible (Fig. \ref{Fig:twocompprofiles}). 
The location of these profiles is only a fraction of a second of arc ($\sim$0\farcs3) 
away from the center of the BP (within the white contours in Fig. \ref{Fig:twocompmaps}),
at the border of the BP (see lower right panel in Fig. \ref{Fig:twocompmaps}). It is 
worth to mention that having a spatial variation of the Stokes $V$ asymmetries within the
magnetic concentration indicates that the internal structure of the BP is at least 
partially resolved.  However, we do not find single-lobed Stokes $V$ profiles as shown 
in the SUNRISE high-resolution observations of an individual network patch
\citep{martinezgonzalez2012}. 
An increment of the asymmetries in area and amplitude of Stokes $V$, away from the core of the BP, 
was interpreted by \citet{riethmuller14} as an indication of a flux-tube canopy. 
Why are the clearly-discernible two-component Stokes $V$
profiles not found at the center of the BP? 
A plausible explanation
is that at the center of the BP, the slower component
is much stronger than the redshifted one. 
Hence, it dominates over the weaker component, which 
still leaves an imprint on the profiles (Fig. \ref{Fig:stokes_BPs}). 
The inferred filling factor from the dual magnetic-component inversions further supports this
hypothesis. The slower (and stronger) component had a filling factor of 0.7, whereas the 
weaker component had only 0.3. This unbalance is likely to be
even stronger at the center of the BP. Therefore, we assume that two components are present 
in and around BPs. Both components are possibly coexisting within the same resolution element at
different atmospheric optical depths. 
As a consequence, strong downflows appear in a region of about 
$1\arcsec \times 1\arcsec$ centered at the BP. 
The strong LOS velocities reach up to 6\,km\,s$^{-1}$, which is 
consistent with downflows inferred from other 
photospheric lines in BPs, such as the \ion{Fe}{i} 6302\,\AA\ and 5250\,\AA\ lines \citep{utz14,nagata08}.

This study has been unable to find large Doppler shifts of the \ion{Na}{i} D$_2$ line in the three isolated 
BPs, contrary to the Doppler shifts of up to 7\,km\,s$^{-1}$ found by \citet[][]{jess10} 
in their data using the \ion{Na}{i} D$_1$ line. Nevertheless, these high velocities were not found in all BPs of their
observations. In the present data, the \ion{Si}{i} Stokes $V$ profiles indicate large flows. 
Figure \ref{Fig:Nad2timeevol} illustrates in time the average \ion{Na}{i} D$_2$ intensity profiles at
the three BPs. There is no indication at any time of large Doppler shifts. Several reasons can be ascribed 
for the non detection of fast flows in this line. First, the large flows are best seen in the 
\ion{Si}{i} Stokes $V$ profiles and are only faintly intuited in the red wing of \ion{Si}{i} Stokes $I$.
Hence, its origin is linked to magnetic structures. Unfortunately, no Stokes $V$ data from \ion{Na}{i} D$_2$ was recorded. 
Therefore, we cannot confirm weather these flows would have left an imprint in the \ion{Na}{i} D$_2$ polarimetric signals.   
Moreover, the low filling factor inferred from the SIR inversions of the fast component (only 0.3), already
indicates that this component contributes less than the slower component. 
Second, the \ion{Na}{i} D$_2$ line is a broader line and a second, weaker component would unlikely  
contribute significantly to the intensity profile. 
It should be also mentioned that, in general, observations are affected by stray light, which is also the case for GREGOR 
\citep[e.g.,][]{borrero16, felipe16, denker18a}. This can affect the apparent 
size of the BP by mainly smearing effects. 
Still, it seems that stray-light contamination does not modify the shape of the Stokes profiles, although it can 
weaken the polarization signals \citep{quinteronoda15}. 
Thus, it is not expected that the present inferred Doppler shifts are significantly influenced by the presence of stray light.

Of particular interest is the case of a forming BP described by \citet{nagata08} 
who deduced a sudden increment of the LOS velocity of $\sim$6\,km\,s$^{-1}$ (downward),
co-temporal with a significant increment of the longitudinal magnetic field. However, 
the shape of the \ion{Fe}{i} 6302\,\AA\ Stokes $I$ profiles from their Fig.~2, 
closely resembles the intensity profiles shown in our Fig. \ref{Fig:stokes_BPs}. 
Likewise, the profiles show asymmetries that are mainly reflected in the red wing, 
which appears inclined toward the red. This signature is seen before, during, and after the sudden 
rise of the longitudinal magnetic field. Therefore, this indicates 
that more than one unresolved component was present in their data all the time within the same resolution element.
Thus, the velocities may have been underestimated before and after the sudden increase of 
the magnetic field. 

In the present study, we lack temporal coverage of the near-infrared Stokes profiles. However, 
we find that all three analyzed BPs, independently of the stage of their life, showed asymmetries 
in the red wing. This suggests that this phenomenon is rather common, 
and likely persistent, during the lifetime of BPs. However, more cases of isolated BPs need to 
be studied. Broader red wings in Stokes~$V$ were attributed to 
downward accelerated and magnetized plasma \citep{steiner98}. 
Such downflows also match with the scenario of convective collapse \citep{parker78} and were
already detected in BPs in the upper photosphere using \ion{Mg}{i}~b$_2$ dopplergrams \citep{fischer09}.
Higher spatial- and spectral-resolution observations will potentially show more cases of strong downflows
associated with BPs.

\subsection{Magnetic field expansion with height}
It is known that the photospheric magnetic field of small- and large-scale
structures expands with height \citep[e.g.,][]{spruit76, solanki93, solanki99, buehler15}. 
Such an expansion happens because gas pressure decreases with height and 
magnetic flux needs to be conserved. 
For instance,  
comparing the sizes of BPs by using filtergrams of two photospheric lines, which are 
sensitive to different optical depths, \citet{rimmele04} deduced that the magnetic flux concentration 
expands with height. \citet{martinezgonzalez2012} studied the asymmetries of \mbox{Stokes $V$} 
profiles in a resolved magnetic patch observed with the balloon-borne observatory \mbox{SUNRISE} and suggested a geometry 
compatible with magnetic features that expand with height. 
Moreover, simulations showed this expansion by illustrating individual 
field lines arising from
flux sheets in intergranular lanes \citep{steiner98}. Here, we search for evidence
of such an expansion in observational data of BPs by analyzing different spectral lines and
carrying out height-dependent inversions. 

First, the circular polarization signals were compared between the
\ion{Ca}{i} 10839\,\AA\  and \ion{Si}{i} 10827\,\AA\ lines.
The former is sensitive to a thin layer in the deep photosphere, while the latter is
formed in the upper photosphere. \citet{felipe18} estimated a difference of about 300\,km in
geometrical height between these two lines in the umbra of a sunspot. Likewise, we assume this as an acceptable
approximation of the height difference for BPs, which are located deeper along dark intergranular lanes. 
The spatial extension across the FOV of the Stokes $V$ signals was about a factor 
of three larger for the \ion{Si}{i} line compared to the \ion{Ca}{i} line (middle panels of Fig.~\ref{Fig:twocompmaps}). 
This difference indicates that the field necessarily spatially expands with height, which is expected
because of the decrease of gas pressure.

Secondly, we scrutinized the \ion{Si}{i} line itself, which covers several optical depths in the solar
atmosphere. To that end, the extension of the inferred longitudinal
magnetic field was compared at two optical depths, $\log \tau = -1$ and $-2.5$. 
All analyzed BPs showed a spatial increment of $B_\mathrm{LOS}$ of at least 7\%. 
A huge rise in extension of about 51\% was deduced for BP (b). 
This disparity in the percentages is owing to the size of the BPs. Figure \ref{Fig:HiFI_zoom} shows 
that all analyzed BPs have a different shape and proportion.
Smaller or individual BPs expand less with height.
Therefore, it is not surprising that the expansion percentages are different for each BP.  

We found rather low values for the longitudinal magnetic fields in BPs, on average 220--340\,G. For individual 
pixels a field strength of up to 620\,G was inferred. We note that these values were computed using a
filling factor that was determined during the inversions (Eq. \ref{Eq:1}). 
The average filling factor in BPs 1--3 was in the range of 0.30 and 0.50. Compared to lower
resolution observations, for instance at the VTT, \citet{beck07} inferred values of $<0.15$. 
This is consistent, since higher spatial resolution observations bring along
larger filling factors.
Furthermore, no linear polarization information was used because the Stokes $Q$ and $U$ signals were 
at the noise level. We speculate that the transverse magnetic
fields might be very small and do not contribute significantly to the magnetic field strength.
From simulations, BPs harbor kilogauss magnetic fields \citep{criscuoli14}. 
The \ion{Si}{i} line is more sensitive to higher layers of the photosphere. 
The expansion of the magnetic fields with height necessarily implies a decrease of the
field strength. We conjecture that this could be the reason for the lower inferred 
fields. We note also that stray light can weaken the polarization signals \citep{quinteronoda15},
thus reducing the inferred field strength.

\section{Summary and conclusions} \label{Sect:6}
The present work aimed at analyzing the magnetic field expansion with height and 
the dynamics of three isolated bright points using high-resolution filtergrams and 
spectropolarimetric data in various spectral lines observed with the
1.5-meter GREGOR solar telescope. To the best of our knowledge, 
the near-infrared spectral range around the \ion{Si}{i} 10827\,\AA\ line has not yet
been explored in relation to BPs. High-resolution filtergrams acquired 
with the HiFI instrument in broad-band \ion{Ca}{ii} H and blue continuum (4505\,\AA)  
were used to easily identify the BPs. Then, spectroscopic images
along the \ion{Na}{i} D$_2$ line and one large spectropolarimetric 
raster scan in the \ion{Si}{i} 10827\,\AA\ spectral range were used to infer 
Doppler velocities and LOS magnetic fields. The observations were taken close to 
disk center. Therefore, we do not expect significant changes between the LOS and 
local solar reference frames. 

The high spatial-resolution filtergrams in the blue-continuum and in the 
\ion{Ca}{ii} H line uncover BPs as thin ($\sim$0\farcs30--0\farcs35) features 
which appear in groups or chains and therefore resemble elongated (few seconds of arc), 
rather than roundish, structures.
The analyzed isolated BPs were visible during the 30-minute observations. 

We summarize the most important findings below:

\begin{itemize}
  \item The three analyzed isolated BPs were identified simultaneously in \ion{Ca}{ii} H (3968\,\AA) and 
        blue continuum (4505\,\AA) filtergrams. Furthermore, they were best detected 
        in filtergrams in the blue wing of the \ion{Na}{i}~D$_2$ line (at $\lambda_0-0.22$\,\AA) and 
        slit-reconstructed images of the \ion{Si}{i} 10827\,\AA\ line (at $\lambda_0-0.35$\,\AA), 
        as well as in the line core of the
        \ion{Ca}{i} 10839\,\AA\ line. 
 
  \item The studied BPs were of magnetic nature, as depicted in Fig.~\ref{Fig:inversion_maps}.
       The longitudinal magnetic fields are on average in the range of 220--340\,G, 
       with individual pixels of up to 620\,G. The fields might be slightly underestimated due to the
       presence of stray light in the observations. 
       
  \item By analyzing the spatial extension of Stokes $V$ signals between the \ion{Ca}{i} and the \ion{Si}{i} lines
        ($\sim$300\,km height difference), we provided evidence for the expansion of the magnetic field 
        with height in BPs (Fig.~\ref{Fig:twocompmaps}). The Stokes $V$ signals of \ion{Si}{i} occupied 
        about three times the area of the \ion{Ca}{i} counterpart. More evidence was provided
        by comparing the spatial extension of the longitudinal magnetic field inferred from the \ion{Si}{i} line
        at two different optical depths 
        ($\log \tau = -1.0$ and $-2.5$). The field expands between 7\% and 51\% with height, 
        depending on the size of the BPs \mbox{(Fig.~\ref{Fig:inversion_maps})}.
  \item The LOS velocities deduced from \ion{Na}{i} D$_2$ at $\lambda_0-0.22$\,\AA\ were moderate to low 
       ($\lesssim$1\,km\,s$^{-1}$).
        Conversely, the \ion{Si}{i} Stokes $I$ and $V$ profiles showed asymmetries which were ascribed to 
        gradients along the LOS (Fig.~\ref{Fig:stokes_BPs}). Some of the Stokes $V$ profiles, about 0\farcs3 away
        from the center of the BP, revealed two discernible components. Two-component
        inversions unveiled downflows of up to 6\,km\,s$^{-1}$ in one component, while the slower one was slightly 
        blueshifted but had a much larger filling factor (Fig.~\ref{Fig:twocompprofiles}). We speculate that both components are also 
        present at the center of the BP but are masked by the stronger (and slower) component. 
        Asymmetries in the red wing of the Stokes profiles
        support this argument (Fig.~\ref{Fig:stokes_BPs}). We suggest that more than one component within the
        resolution element may be common in BPs. 
  \item The \ion{He}{i} 10830\,\AA\ triplet was not sensitive to the studied isolated BPs. However, 
        we cannot rule out weak polarization signals which may become visible when using larger exposure
        times. 
\end{itemize}

Our findings shed new light on the dynamics and the magnetic field expansion present in BPs. 
Further work needs to be carried out to follow the time evolution of Stokes profiles in BPs with high 
temporal, spatial, and spectral resolution. An important role will play the multi-wavelength capabilities
of the next generation of large-aperture telescopes. Such multi-line observations are
crucial to better follow the magnetic field lines and flows across the solar atmosphere.  
We expect further advances with the advent of the 4-m Daniel K. Inouye Solar Telescope 
\citep[DKIST,][]{dkist} and European Solar Telescope \citep[EST,][]{est}.  

\begin{acknowledgements}
The 1.5-meter GREGOR solar telescope was built by a German consortium under the leadership of the 
Leibniz-Institut f\"ur Sonnenphysik in Freiburg (KIS) with the Leibniz-Institut f\"ur Astrophysik Potsdam (AIP), 
the Institut f\"ur Astrophysik G\"ottingen (IAG), the Max-Planck-Institut f\"ur Sonnensystemforschung in 
G\"ottingen (MPS), and the Instituto de Astrof\'isica de Canarias (IAC), and with contributions by the Astronomical 
Institute of the Academy of Sciences of the Czech Republic (ASCR).
The author thanks Drs. H. Balthasar and S. J. Gonz\'alez Manrique for their help 
during the observations. Drs. H. Balthasar, S. J. Gonz\'alez Manrique, I. Kontogiannis, and D. Utz, 
are greatly acknowledged for helpful discussions and for providing very valuable input on the manuscript. 
In addition, the author thanks Dr. C. Denker 
for carefully reading the manuscript and providing suggestions, as well as the anonymous referees for their
criticism which led to an improved version of the manuscript. 
This research has made use of NASA's Astrophysics Data System. Funding from the Horizon 2020 projects 
PRE-EST (grant agreement No 739500) and SOLARNET (No 824135) is greatly acknowledged.

\end{acknowledgements}


\bibliographystyle{aa}
\bibliography{aa-jour,biblio}

\end{document}